\documentclass[apjs]{emulateapj}

\shorttitle{IFS of 23 Spiral Bulges}
\shortauthors{Batcheldor et al.}

\begin{document}

\title{Integral Field Spectroscopy of 23 Spiral Bulges}

\author{D. Batcheldor,\altaffilmark{1,2} D. Axon,\altaffilmark{2,1} D. Merritt,\altaffilmark{2} M. A. Hughes,\altaffilmark{1} 
A. Marconi,\altaffilmark{3} J. Binney,\altaffilmark{4} A. Capetti,\altaffilmark{5} M. Merrifield,\altaffilmark{6} 
C. Scarlata,\altaffilmark{7} and W. Sparks\altaffilmark{8}}
\journalinfo{The Astrophysical Journal Supplement Series}
\submitted{Received 2005 March 17; accepted 2005 April 27}

\altaffiltext{1}{Centre for Astrophysics Research, STRI, University of Hertfordshire, Hatfield, AL10 9AB, UK; danb@star.herts.ac.uk}
\altaffiltext{2}{Department of Physics, Rochester Institute of Technology, 84 Lomb Memorial Drive, Rochester, NY, 14623, USA.}
\altaffiltext{3}{INAF-Osservatorio Astrofisico di Arcetri, Largo E. Fermi 5, 50125 Firenze, Italy.}
\altaffiltext{4}{Oxford University, Theoretical Physics, Keble Road, Oxford, OX1 3NP, UK.}
\altaffiltext{5}{INAF-Osservatorio Astronomico di Torino, I-10025 Pino Torinese, Italy.}
\altaffiltext{6}{School of Physics and Astronomy, University of Nottingham, NG7 2RD, UK.}
\altaffiltext{7}{Eidgenoessische Technische Hochshule Z\"{u}rich, H\"{o}nggerberg HPF G4.3, CH-8092 Z\"{u}rich, Switzerland.}
\altaffiltext{8}{Space Telescope Science Institute, 3700 San Martin Drive, Baltimore, MD, 21218, USA.}

\begin{abstract}
We have obtained Integral Field Spectroscopy for 23 spiral bulges using INTEGRAL on the William Herschel Telescope and SPIRAL on the 
Anglo-Australian Telescope. This is the first 2D survey directed solely at the bulges of spiral galaxies. Eleven galaxies of the sample 
do not have previous measurements of the stellar velocity dispersion ($\sigma_{\ast}$). These data are designed to complement our Space 
Telescope Imaging Spectrograph program for estimating black hole masses in the range $10^{6}-10^{8}M_{\sun}$ using gas kinematics from 
nucleated disks. These observations will serve to derive the stellar dynamical bulge properties using the traditional Mgb and CaII 
triplets. We use both Cross Correlation and Maximum Penalized Likelihood to determine projected $\sigma_{\ast}$ in these systems and 
present radial velocity fields, major axis rotation curves, curves of growth and $\sigma_{\ast}$ fields. Using the Cross Correlation to 
extract the low order 2D stellar dynamics we generally see coherent radial rotation and irregular velocity dispersion fields suggesting 
that $\sigma_{\ast}$ is a non-trivial parameter to estimate.
\end{abstract}


\section{Introduction}

The increasing number of super-massive black hole (SMBH) candidates discovered in the mid 1990s led to the investigation of SMBH 
demographics. It was found that various properties of the host bulge correlated with the SMBH mass, e.g. bulge luminosity and mass 
\citep{kandr95,mag98}, stellar velocity dispersion - the $M_{\bullet}-\sigma_{\ast}$ relation - \citep{fandm00,geb00} and concentration 
index \citep{gra03}. Scatter in these relations are comparable if enough care is taken in deriving the bulge properties \citep{mandh03}. 
Such correlations have fundamental implications for both SMBH and host galaxy formation and evolution. These ``secondary'' methods also 
present a valuable practical tool to estimate SMBH mass. The ``primary'' methods, used to originally define the relations, determine 
SMBH mass by considering the direct gravitational effects of SMBHs. However, each primary method has a drawback: proper motion studies 
can only be carried out in the Milky Way; ${\rm H_{2}0}$ masers need to be orientated correctly; gas kinematics may be influenced by 
non-gravitational effects and stellar dynamics must be derived from relatively low surface brightnesses. The advantages of knowing the 
exact form of the SMBH - host relations then become obvious if trying to sample a large number of SMBH masses across a wide dichotomy of 
host types and redshifts.

\begin{figure*}
\plotone{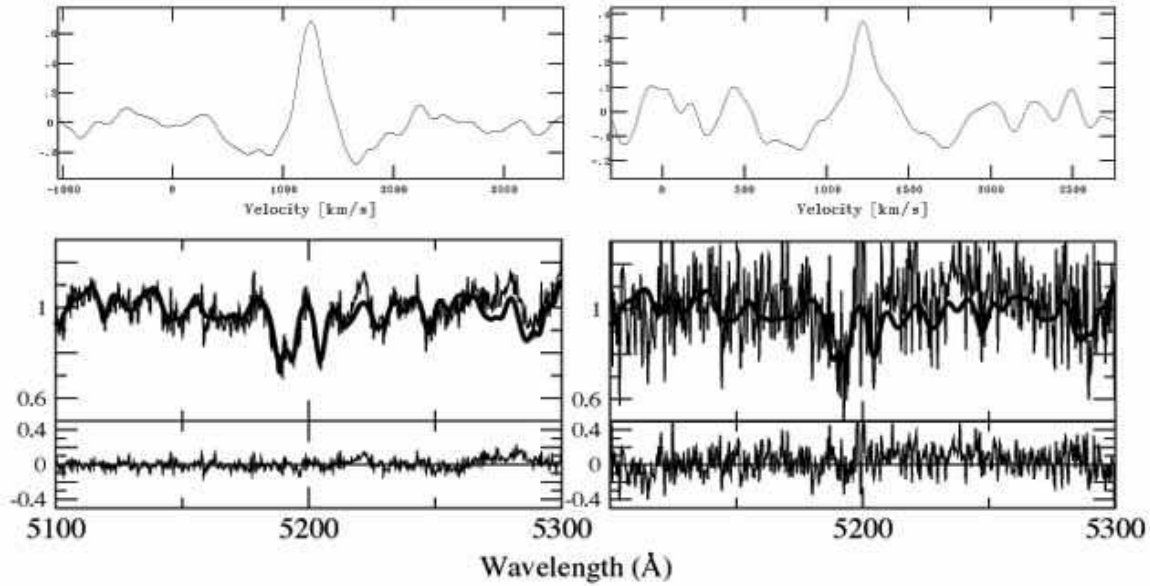}
\caption[Xcor Example]{
Demonstrating the Xcor fits to the data. [top row] The left panel shows the cross correlation function derived from a S/N=23 spectrum in 
NGC~4041, while the right panel details the function in a S/N=8 case. [bottom row] The bold lines show the fits using the derived 
parameters to the observed spectra. The lower sections present the residuals of the fits. 
}\label{fig:4041example}
\end{figure*}

There still exists a fundamental bias toward large ($\sim 10^{9}M_{\sun}$) SMBH mass measurements in the relations however. This 
stems from the relative ease of mass measurements in giant ellipticals: it is easier to spot a bigger black hole. The low mass end 
of the relations ($10^{6}M_{\sun} < M_{\bullet} < 10^{8}M_{\sun}$), which essentially represents late type spirals, has yet to be 
fully populated. This is unsurprising when considering the difficulties and complexities involved with observing and modelling such 
systems. This swing toward high mass SMBHs has led some to discuss whether these relations are indeed linear \citep{laor01}.

If we are to effectively extrapolate these secondary relations to make conclusions about formation and evolutionary scenarios, and 
indeed use them as SMBH mass estimators, we need to quell any doubts as to their nature. To this end we have completed a 
Space Telescope Imaging Spectrograph (STIS) survey of 54 local spirals \citep{hugh03,scar04} using the {\it Hubble Space Telescope} 
({\it HST}). This data has been used to make primary SMBH mass estimates by modelling the gravitational potential as represented by 
the central gas kinematics and surface brightness profiles \citep{mar03,atk04}. However, to populate the low mass end of the 
secondary estimators we need to establish the bulge properties. Consequently we have initiated integral field spectroscopic observations 
of the original STIS sample. Integral Field Spectroscopy (IFS) is a reliable and highly flexible tool capable of quickly gathering a 
large amount of dynamical information. It allows the observation of the full 2D Line of Sight Velocity Distributions (LOSVDs). 

As there are few published 2D kinematical maps of spiral bulges we present here an atlas of our data so far including radial velocity 
fields, major axis rotation curves, stellar velocity dispersion ($\sigma_{\ast}$) fields and aperture integrated velocity dispersions 
(curves of growth). In \S~\ref{obs+dr} we outline the observations and basic data reduction, while \S~\ref{losvd} details how we extracted 
the 2D LOSVDs. The data is presented in \S~\ref{data} and briefly summarized and discussed in \S~\ref{summary}.

\begin{figure*}
\plotone{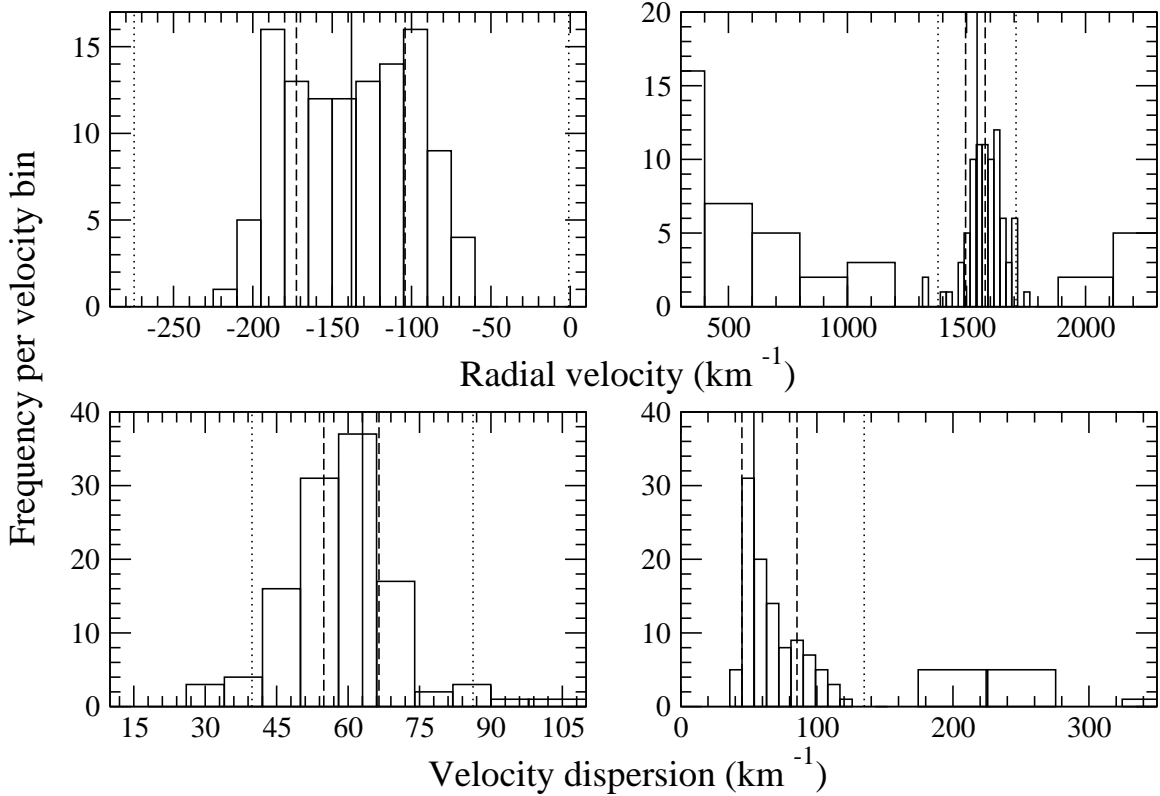}
\caption[Discordance Example]{
Identifying discordant data. In all cases the solid line marks the median value and the dashed lines show the upper and lower quartiles. 
The dotted lines mark the $2i\pm m$ boundaries, where $m$ is the median and $i$ is the inter-quartile range. Data falling outside of the 
dotted lines were marked as discordant. [left column] NGC~4212. The top panel presents the radial velocities and the bottom panel shows 
the velocity dispersions in this low discordant example. [right column] Same as the left column but for NGC~3021; a high discordant 
example.
}\label{fig:discord}
\end{figure*}

\section{Observations and Data Reduction}\label{obs+dr}

Our observations form the first IFS survey specifically applied to a sample of spiral bulges. They positively complement the SAURON 
IFS project which is directed at 72 targets including a number of spirals \citep{dez02,ems04}. From the nature of the 
$M_{\bullet}-\sigma_{\ast}$ relation, and the general character of spiral galaxies, we required instruments capable of resolving 
$\sigma_{\ast}$ well below $100{\rm~km~s^{-1}}$. For our northern hemisphere targets this lead us to use the fibre fed Integral Field 
Unit (IFU) INTEGRAL, on the 4.2m WHT, with the $5^{\rm th}$ order Echelle grating ($\Re\approx6000$). Whilst the CaII triplet is the 
preferred tracer of stellar dynamics \citep{dre84} this grating could only be centered on the Mgb triplet ($\sim5200{\rm\AA}$). 
Observing conditions ensured all nights were non-photometric and forced the use of the SB3 fibre bundle. This bundle consists of a main 
$33\farcs6\times 29\farcs4$ array with 115 circular apertures\footnote{Throughout ``apertures'' means the individual fibres, 
lenslets or spaxels referred to in other IFS studies.} packed according to Kepler's Conjecture. The $2\farcs7$ diameter of these 
apertures ensured that the PSF due to seeing (which was typically $1\farcs5$) was under-sampled. The main array is surrounded by a 
concentric circle ($\diameter90\arcsec$) of 20 apertures designed to sample the sky. For our southern hemisphere targets we used the 
fibre fed IFU SPIRAL, on the 3.9m AAT, with the R1200 grating ($\Re\approx7600$) centered on the CaII triplet at $\sim8660{\rm\AA}$. 
Ordinarily SPIRAL consists of 512 apertures but effective sky reduction can be achieved by masking half of the array and using SPIRAL 
in ``nod and shuffle'' (NS) mode. NS masking of the MITLL2A detector meant that 1 row and 1 column of the array fell off the chip. This 
reduced the number of usable apertures ($\diameter0\farcs7$) to 225 over a $11\arcsec\times9\farcs8$ area. 

During our observing runs priority was given to targets for which we have successful STIS spectra \citep{hugh03}. To date we have 
completed IFS observations on 23 of the original 54 spiral bulges. Calibration frames including flat-fields, tungsten lamp flats, dark 
frames, bias frames and Cu-Ne/Cu-Ar arc frames, along with at least 3 late type (G, K, M) spectroscopic standards, were observed 
throughout each night. At least 3 target exposures were made in order to facilitate the easy removal of cosmic rays. We made several 
offset sky observations, especially when the overheads associated with NS observations lead us to abandon its use in some cases. The 
SPIRAL data was reduced using the adapted automatic 2df software developed by Jeremy Bailey of the Anglo-Australian Observatory. INTEGRAL 
data reduction was carried out with standard IRAF\footnote{IRAF is distributed by the National Optical Astronomy Observatories, which are 
operated by the Association of Universities for Research in Astronomy, Inc., under co-operative agreement with the National Science 
foundation.} routines, and made use of the {\it integral} package developed by Robert Greimel of the Isaac Newton Group. Detector 
positions of the individual aperture spectra were traced by fitting a polynomial across the dispersion axis of a high signal lamp flat. 
This enabled scattered light to be identified and removed from the science frames. The trace of the spectra was then used as a master 
template in extracting the subsequent spectra. Lamp flats were again used to make throughput corrections before carrying out wavelength 
calibrations. Sky subtraction was performed using NS or by subtracting a collapsed off-set sky frame or summed sky apertures. All objects 
had the continuum fitted to each individual spectra. This level was then used together with the aperture spacial positions to create 2D 
intensity maps (essentially very low spacial resolution images of the targets). 

\section{Extracting LOSVDs}\label{losvd}

\begin{deluxetable*}{ccccccccc}
\tabletypesize{\footnotesize}
\tablecaption{Main Results for the Sample \label{tab:results}}
\tablewidth{0pt}
\tablehead{
\colhead{}                  &\colhead{}                  &\colhead{} & \colhead{$\sigma_{\rm ce}$} & \colhead{$\sigma_{\rm cc}$} & 
\colhead{$\sigma_{\rm me}$} &\colhead{$\sigma_{\rm mc}$} &\colhead{} \\
\colhead{Galaxy}            &\colhead{\% DC}             & Peak S/N                   &\colhead{${\rm(km~s^{-1})}$}&
\colhead{${\rm(km~s^{-1})}$}&\colhead{${\rm(km~s^{-1})}$}&\colhead{${\rm(km~s^{-1})}$}&\colhead{$h_3$}             &\colhead{$h_4$} \\
}
\startdata
\object{NGC~0289} & 30.4 & 19.69 &$95\pm10$  &$114\pm14$&$93^{+41}_{-29}$  &$111^{+47}_{-24}$ &$0.02^{+0.01}_{-0.05}$ &$0.26^{+0.03}_{-0.02}$ \\
\object{NGC~0613} & 29.6 & 20.70 &$113\pm1$  &$99\pm2$  &$201^{+9}_{-10}$  &$149^{+19}_{-22}$ &$-0.14^{+0.02}_{-0.05}$&$0.11^{+0.08}_{-0.07}$ \\  
\object{NGC~1255} & 16.5 & 10.94 &$57\pm1$   &$69\pm5$  &$34^{+32}_{-9}$   &$47^{+27}_{-3}$   &$-0.14^{+0.09}_{-0.12}$&$0.09^{+0.13}_{-0.03}$ \\  
\object{NGC~1300} & 11.3 & 21.40 &$87\pm5$   &$90\pm6$  &$86^{+32}_{-13}$  &$82^{+55}_{-19}$  &$0.12^{+0.01}_{-0.20}$ &$0.16^{+0.06}_{-0.11}$ \\  
\object{NGC~1832} & 13.0 & 15.98 &$88\pm3$   &$102\pm13$&$104^{+11}_{-13}$ &$129^{+16}_{-18}$ &$-0.09^{+0.09}_{-0.08}$&$0.03^{+0.04}_{-0.04}$ \\  
\object{NGC~2748} & 17.4 & 20.98 &$83\pm8$   &$79\pm7$  &$107^{+7}_{-10}$  &$78^{+3}_{-7}$    &$0.00^{+0.06}_{-0.07}$ &$-0.02^{+0.05}_{-0.05}$\\  
\object{NGC~2903} & 33.0 & 25.84 &$89\pm4$   &$95\pm6$  &$138^{+5}_{-6}$   &$171^{-0}_{-6}$   &$-0.12^{+0.08}_{-0.10}$&$0.15^{+0.09}_{-0.14}$ \\  
\object{NGC~2964} & 19.1 & 29.33 &$103\pm12$ &$111\pm14$&$138^{+28}_{-18}$ &$135^{+40}_{-26}$ &$0.07^{+0.11}_{-0.09}$ &$0.21^{+0.03}_{-0.05}$ \\  
\object{NGC~3021} & 40.0 & 14.54 &$82\pm3$   &$125\pm7$ &$81^{+74}_{-1}$   &$62^{+45}_{-15}$  &$-0.08^{+0.07}_{-0.06}$&$0.12^{+0.15}_{-0.12}$ \\  
\object{NGC~3162} & 24.4 & 14.18 &$70\pm2$   &$85\pm1$  &$80^{+11}_{-10}$  &$82^{+12}_{-7}$   &$-0.12^{+0.04}_{-0.01}$&$0.07^{+0.02}_{-0.04}$ \\  
\object{NGC~3310} & 26.1 & 27.74 &$84\pm1$   &$101\pm1$ &$127^{+25}_{-19}$ &$115^{+15}_{-17}$ &$0.01^{+0.01}_{-0.08}$ &$0.15^{+0.04}_{-0.10}$ \\  
\object{NGC~3949} & 38.3 & 18.95 &$82\pm2$   &$80\pm3$  &$113^{+17}_{-12}$ &$75^{+7}_{-7}$    &$0.00^{+0.03}_{-0.05}$ &$0.21^{+0.03}_{-0.06}$ \\  
\object{NGC~3953} & 12.2 & 19.95 &$116\pm3$  &$146\pm1$ &$122^{+4}_{-8}$   &$127^{+13}_{-12}$ &$0.02^{+0.01}_{-0.02}$ &$-0.01^{+0.04}_{-0.01}$\\  
\object{NGC~4041} & 11.3 & 22.93 &$88\pm7$   &$92\pm8$  &$97^{+7}_{-3}$    &$96^{+5}_{-6}$    &$0.04^{+0.06}_{-0.04}$ &$0.00^{+0.04}_{-0.04}$ \\  
\object{NGC~4051} & 40.0 & 45.05 &$72\pm10$  &$91\pm10$ &$92^{+19}_{-15}$  &$126^{+30}_{-20}$ &$1.7^{+0.5}_{-0.9}$    &$2.1^{+0.3}_{-0.6}$    \\  
\object{NGC~4088} & 27.0 & 15.46 &$77\pm2$   &$87\pm4$  &$88^{+8}_{-9}$    &$117^{+25}_{-18}$ &$0.00^{+0.04}_{-0.04}$ &$0.11^{+0.04}_{-0.03}$ \\  
\object{NGC~4212} &  4.4 & 21.05 &$75\pm2$   &$82\pm5$  &$62^{+3}_{-4}$    &$67^{+6}_{-4}$    &$-0.02^{+0.14}_{-0.10}$&$0.07^{+0.04}_{-0.11}$ \\  
\object{NGC~4258} & 20.9 & 39.38 &$148\pm4$  &$120\pm6$ &$126^{+9}_{-4}$   &$140^{+12}_{-6}$  &$-0.01^{+0.01}_{-0.02}$&$-0.02^{+0.01}_{-0.01}$\\  
\object{NGC~4303} & 23.5 & 29.71 &$84\pm3$   &$108\pm2$ &$102^{+2}_{-6}$   &$133^{+10}_{-22}$ &$-0.06^{+0.04}_{-0.03}$&$0.01^{+0.02}_{-0.03}$ \\  
\object{NGC~4321} & 16.5 & 21.33 &$83\pm3$   &$101\pm2$ &$95^{+7}_{-5}$    &$109^{+6}_{-16}$  &$-0.07^{+0.04}_{-0.03}$&$0.11^{+0.03}_{-0.02}$ \\  
\object{NGC~4536} & 37.4 & 14.62 &$85\pm1$   &$168\pm13$&$182^{+16}_{-20}$ &$190^{+7}_{-24}$  &$0.01^{+0.08}_{-0.04}$ &$0.02^{+0.07}_{-0.07}$ \\  
\object{NGC~5005} & 10.4 & 53.71 &$154\pm10$ &$203\pm11$&$190^{+7}_{-6}$   &$213^{+7}_{-10}$  &$0.06^{+0.04}_{-0.03}$ &$0.05^{+0.03}_{-0.02}$ \\  
\object{NGC~5055} & 12.2 & 47.12 &$117\pm6$  &$118\pm5$ &$124^{+5}_{-4}$   &$122^{+7}_{-4}$   &$0.04^{+0.02}_{-0.02}$ &$0.04^{+0.03}_{-0.01}$ \\
\enddata
\tablecomments{``\% DC'' refers to the percentage of discordant data. ``$\sigma_{\rm ce}$'' indicates extended Xcor values, 
``$\sigma_{\rm cc}$'' central Xcor, ``$\sigma_{\rm me}$'' extended MPL and ``$\sigma_{\rm mc}$'' central MPL. ``$h_3$'' and ``$h_4$'' 
are for the higher LOSVD moments in the central apertures.}
\end{deluxetable*}

Since the early 1970s there have been a multitude of techniques employed to extract dynamical information from digitized spectra 
\citep{sim74,sarg77,fih89,bend90,randw92,vdmandf93,kandm93,merr97,capp04}. In general each method makes the fundamental assumption that 
LOSVDs can be recovered by convolving a broadening function to a shifted template stellar spectra of the same mean spectral type. We 
have attempted to use each of these methods on our sample, but only a select few are suited to the nature of our data (essentially 
$\sim3500$ S/N$\approx10$ spectra). Typical Fourier and pixel fitting techniques fail at the levels of S/N present here and require a 
large amount of time to complete due to the nature of filters and number of possible solutions. We have seen that the technique of Cross 
Correlation, hereafter Xcor \citep[TD79]{tandd79}, performed well in extracting accurate $1^{\rm st}$ and $2^{\rm nd}$ order LOSVD 
moments from our spectra and is particularly suited to large datasets.

We performed Xcor on each spectrum for each observed template by following TD79 as implemented in the {\it iraf.rvsao} package. As these 
reduction steps are considered standard and well established we refer readers wishing to see the detailed treatment of spectra through 
the cross correlation procedure to \citet{kandm98}. In brief, the position of the Xcor peak determines the radial velocity shift, whilst 
the width of the peak is related to the projected stellar velocity dispersion ($\sigma_{\ast}$). Each spectrum has any residual emission 
features suppressed and is continuum subtracted, apodized, zero-padded and binned linearly in $\log$ space before being passed through 
the Xcor. Quadratic weights are applied to deviations in the continuum subtracted spectra to ensure that stronger lines are more heavily 
weighted. To limit the size of secondary peaks in the Xcor function a cosine bell filter is applied. 

Errors in the position and width of the Xcor are defined by the ``$r$'' statistic which is calculated by {\it rvsao} and is used as a 
quality measure of the Xcor. If we assume that a perfect Xcor is proportional to the broadening function convolved with the Auto 
Correlation Function (ACF, i.e., the Xcor of a spectrum with itself) then we expect to see symmetry about the radial velocity. This is 
because both the broadening function and the ACF are themselves symmetrical. The constant of proportionality between the perfect 
correlation and the convolved broadening function will be defined by the height of the central peak. The height of the peak, normalized 
by the antisymmetric components in the Xcor, defines $r$ and is used as the aforementioned quality measure of values derived from the 
Xcor itself. 

However, these error estimates are insignificant when compared to those introduced by the possibility of template mismatch, low signal 
spectra and large velocity dispersions. The advantages of observing a template library now becomes apparent. For each aperture, after 
correcting the Xcor radial velocities for barycentric and stellar radial velocities, we used the $r$ weighted average value between each 
template for the velocity and width measured in that aperture. The standard deviation around these weighted values was then used as a 
separate estimate in the error of each measurement. More weight was placed on higher $r$ values as defined in the following equation:

\begin{equation}\label{equ:weight}
\bar{\phi_F} = \frac{\sum^3_{i=1}\phi_i r^2_i}{\sum^3_{i=1}r^2_i} 
\end{equation}

where $\phi_F$ is the final velocity or width and $\phi_{1..3}$ and $r_{1..3}$ represent the values derived using template 1,2 or 3. Any 
of the weighted values that fell outside the equivalent of a $2.7\sigma$ clip (defined by the interquartile range of the data from the 
entire array) were flagged as ``discordant'', the percentage of discordant events per galaxy then being used as a quality measure of the 
IFS data. 

As with the intensity maps, the full 2D kinematics were recovered by mapping the derived velocities to the relative spacial positions of 
each aperture on the IFUs.

We have carried out extensive testing of the Xcor procedure by creating a composite spectra from the observed templates, generating 
simulated radial velocity and velocity dispersion fields, introducing a surface brightness profile and adding Poisson noise. As the 
width of the Xcor peak does not give a direct measure of $\sigma_{\ast}$ we tested three independent $\sigma_{\ast}$ recovery methods: 
HWHM correction ($\sigma\sqrt{2\ln{2}}$); ACF correction ($\sigma^2 = \mu^2 - 2\tau^2$, where $\mu$ is the width of the peak and $\tau$ 
is given by the ACF) and a calibration function correction. The calibration function, which was determined by fitting a polynomial to the 
known and measured values of $\sigma_{\ast}$, proved to be far superior in terms of $\chi^2$. As some of our sample are known to contain 
starbursts and AGN we also tested the effects that such objects may have on the Xcor. To the composite template spectrum we added 
increasing proportions of early type spectra and a Seyfert 1 spectrum. In both cases we found that 90\% contamination was needed in order 
to significantly affect the $r$ statistic (i.e., take it below 3). Similar results were also found when introducing large velocity shifts 
to spectra ($\sim3000{\rm~km~s^{-1}}$).

Examples of the Xcor are presented in Figures~\ref{fig:4041example} and \ref{fig:discord}. Figure~\ref{fig:4041example} shows the cross 
correlation functions from two apertures in NGC~4041 and overlays of the derived fits onto observed spectra. In both the high signal 
(S/N=23) and low signal (S/N=8) cases well defined cross correlation functions can be seen giving confident results. These Figures also 
demonstrate the nature of the $r$ statistic (the amount of anti-symmetry in the cross correlation function). Figure~\ref{fig:discord} 
demonstrates how discordant data was identified and flagged. The two cases presented show both low and high discordance observations. 
The derived radial velocity and velocity dispersion data from low discordance and high discordance cases (NGC~4212 and NGC~3021 
respectively) were placed into discrete bins in velocity space. The distribution of data is shown by plotting the number of measurements 
to fall within a specific bin.

\begin{figure}
\epsscale{1.0}
\plotone{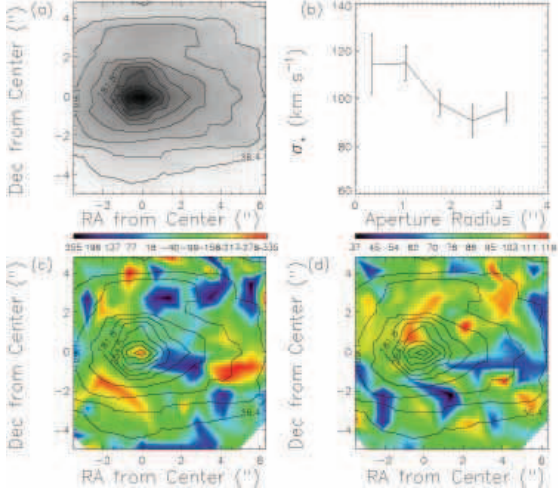}
\caption[NGC~0289]{
Xcor results for NGC~0289. (a) 2D intensity map. (b) Curve of growth. (c) Radial Velocity Field ${\rm(km~s^{-1})}$. (d) $\sigma_{\ast}$ 
Field ${\rm(km~s^{-1})}$. Full color plots can be found in the electronic version.
}\label{fig:289}
\end{figure}
\begin{figure}
\epsscale{1.0}
\plotone{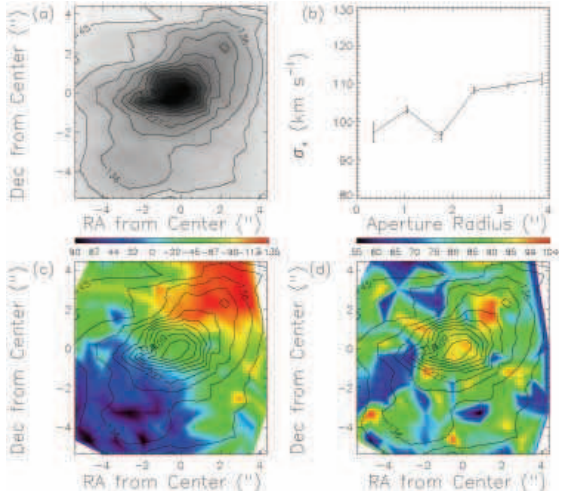}
\caption[NGC~613]{
Same as Figure~\ref{fig:289} but for NGC~0613.
}\label{fig:613}
\end{figure}

In spectra down to S/N$\approx5$ we are confident that the Xcor procedure will return accurate values for the $1^{\rm st}$ and 
$2^{\rm nd}$ order LOSVD moments. However, in many cases the central apertures of the IFU arrays contain S/N $>15$. Here we can attempt 
to recover higher order moments ($h_3$ and $h_4$) which harbor further dynamical information on the observed system, i.e., signatures of 
SMBHs. Performing LOSVD extraction via an independent technique also provides a valuable verification of previous results. The Maximum 
Penalized Likelihood, hereafter MPL \citep{merr97}, is a non-parametric method in that it places no constraints on the shape of the 
LOSVD and allows the recovery of higher order moments ($h_0...h_6$). It is ideally suited to low S/N as the penalty function introduces 
significantly less bias than Fourier methods when smoothing noise characteristics. We carried out similar tests as for the Xcor and 
again found that 90\% starburst/AGN contamination and shifts of $>3000{\rm~km~s^{-1}}$ were needed to significantly affect the 
extracted moments. In order to determine the most appropriate value for the smoothing parameter ($\alpha$) we re-performed the MPL for 
increasing values of $\alpha$ until cusps in the higher moments were seen. A bootstrap, using 100 realizations, was used in order to 
determine the 68\% confidence levels in the derived Hermite polynomials.
\vspace{1cm}

\section{Results, Velocity Fields and Comments for Individual Galaxies}\label{data}

The main results for our sample are presented in Table~\ref{tab:results}. We have provided measures of $\sigma_{\ast}$ taken from the 
central aperture of each array (determined by the peak value of $\sigma_{\ast}$ closest to the photometric center of the target - 
``central'' apertures) and also $\sigma_{\ast}$ measured from spectra collapsed across the entire arrays (``extended'' apertures). 
We also present, for each galaxy, intensity maps derived from the continuum level of each spectra, the ``curve of growth'' derived by
determining $\sigma_{\ast}$ through progressively larger collapsed apertures, the radial velocity field and the velocity dispersion 
field. In all cases North is up and East is to the left. In the cases where there is evidence of a net rotation the radial velocities 
along the major axes have also been presented. We continue by making comments on each galaxy. 

\subsection{NGC~0289}

Figure~\ref{fig:289}. The data are noisy and almost 1/3 is excluded as discordant. As a consequence nothing can be said about the 2D 
dynamics. These are the first measures of $\sigma_{\ast}$ in NGC~289.

\begin{figure}
\epsscale{1.0}
\plotone{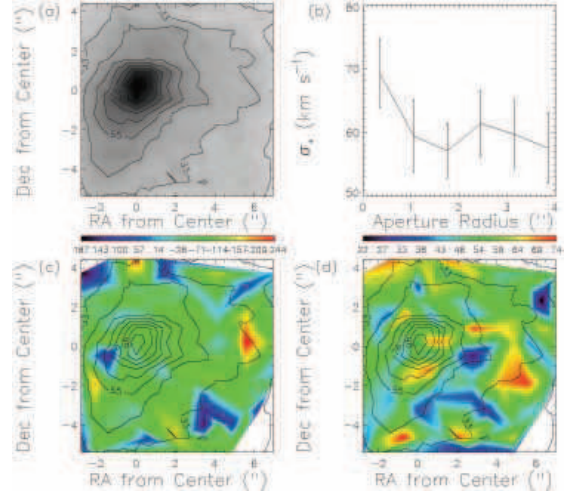}
\caption[NGC~1255]{
Same as Figure~\ref{fig:289} but for NGC~1255.
}\label{fig:1255}
\end{figure}
\begin{figure}
\epsscale{1.0}
\plotone{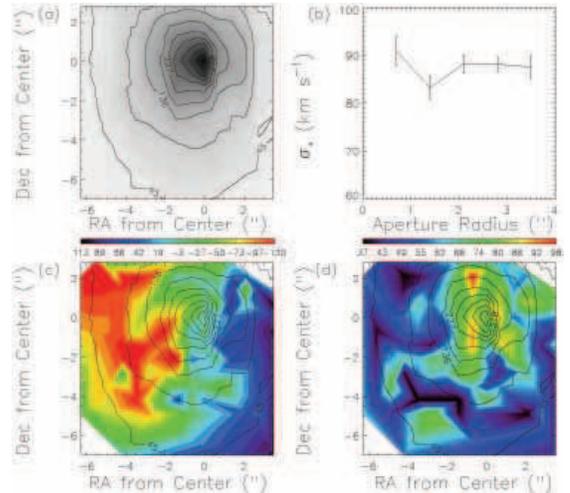}
\caption[NGC~1300]{
Same as Figure~\ref{fig:289} but for NGC~1300.
}\label{fig:1300}
\end{figure}

\subsection{NGC~0613}

Figures~\ref{fig:613} and \ref{fig:rot613-2903}(a). Clearly defined rotation despite a high level of discordance. The $\sigma_{\ast}$ 
field is patchy but a significant difference between extended and central values of $\sigma_{\ast}$ is noted. There are also significant 
values of $h_3$ and $h_4$ corresponding to a central asymmetric non-Gaussian LOSVD.

\subsection{NGC~1255}

Figure~\ref{fig:1255}. S/N is low in all spectra but the level of discordance is also low. No clear rotation can be seen and the 
$\sigma_{\ast}$ field is patchy. These are the first measures of $\sigma_{\ast}$ in NGC~1255.

\subsection{NGC~1300}

Figures~\ref{fig:1300} and \ref{fig:rot613-2903}(b). A low level of discordance and good S/N has produced a velocity field showing 
rotation and a clear centrally peaked $\sigma_{\ast}$ field. Color maps suggest possible obscuration from spiral structures 
\citep{hugh03}. 

\begin{figure}
\epsscale{1.0}
\plotone{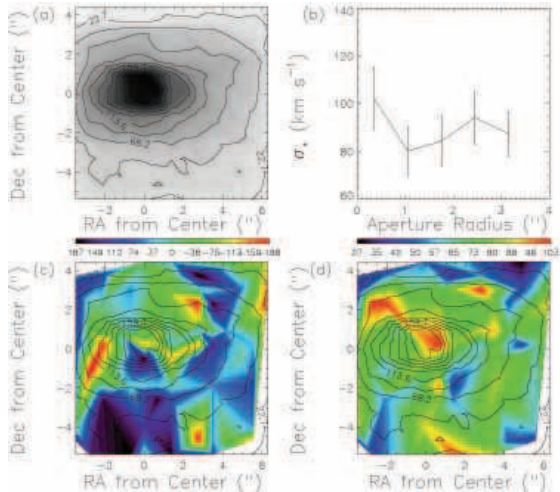}
\caption[NGC~1832]{
Same as Figure~\ref{fig:289} but for NGC~1832.
}\label{fig:1832}
\end{figure}
\begin{figure}
\epsscale{1.0}
\plotone{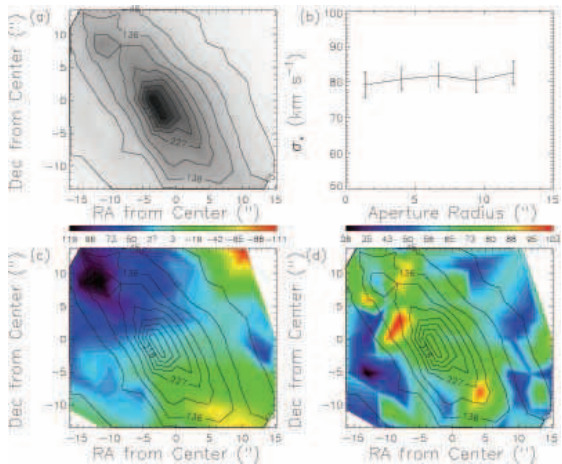}
\caption[NGC~2748]{
Same as Figure~\ref{fig:289} but for NGC~2748.
}\label{fig:2748}
\end{figure}

\subsection{NGC~1832}

Figure~\ref{fig:1832}. The levels of S/N are low and the radial velocity and $\sigma_{\ast}$ fields are noisy. There is a significant 
difference between central and extended values of $\sigma_{\ast}$. These are the first measures of $\sigma_{\ast}$ in NGC~1832.

\subsection{NGC~2748}

Figures~\ref{fig:2748} and \ref{fig:rot613-2903}(c). Typical S/N and discordance. Clear rotation and patchy $\sigma_{\ast}$. NGC~2748 has 
an inclination of $\sim70\degr$ which could account for LOSVD obscuration especially in the presence of dust lanes \citep{hugh03}. These 
are the first measures of $\sigma_{\ast}$ in NGC~2748 and are constant with aperture radius. 

\subsection{NGC~2903}

\begin{figure}
\epsscale{1.0}
\plotone{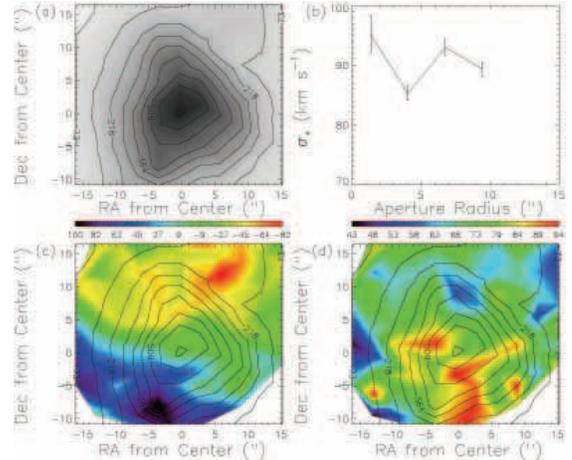}
\caption[NGC~2903]{
Same as Figure~\ref{fig:289} but for NGC~2903.
}\label{fig:2903}
\end{figure}

\begin{figure}
\epsscale{1.0}
\plotone{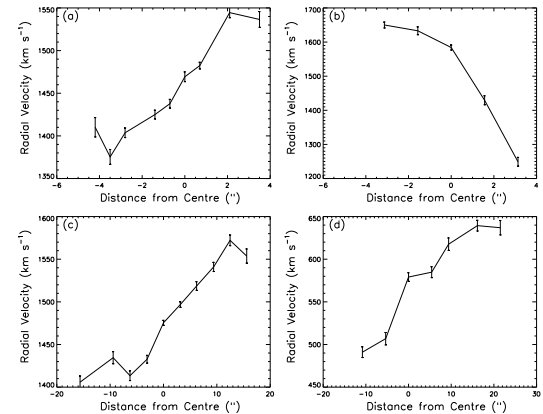}
\caption[Rotation Curves, NGC~613...NGC~2903]{
Radial velocity curves taken along the major axes of each galaxy. Numbers in square brackets refer to the position angle in degrees (east 
from north) along which the curves have been extracted. (a) NGC~613 [135]. (b) NGC~1300 [90]. (c) NGC~2748 [30]. (d) NGC~2903 [150].
}\label{fig:rot613-2903}
\end{figure}

Figures~\ref{fig:2903} and \ref{fig:rot613-2903}(d). High level of discordance but S/N is good. Clear rotation and an irregular 
$\sigma_{\ast}$ field. MPL did not converge in the central aperture. A known star-burst with a circumnuclear ring \citep{ark01}.


\subsection{NGC~2964}

\begin{figure}
\epsscale{1.0}
\plotone{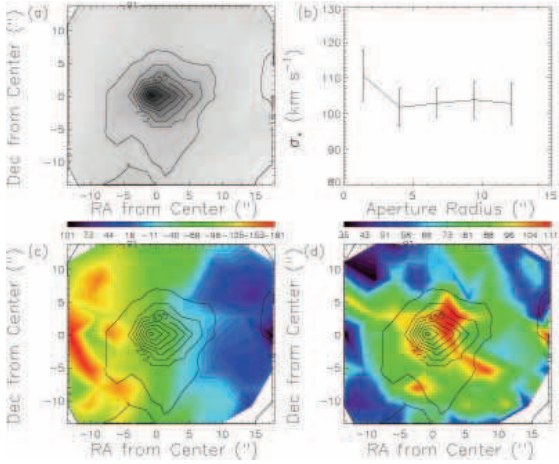}
\caption[NGC~2964]{
Same as Figure~\ref{fig:289} but for NGC~2964.
}\label{fig:2964}
\end{figure}

Figures~\ref{fig:2964} and \ref{fig:rot2964-3310}(a). Good S/N with clear rotation and a centrally peaked $\sigma_{\ast}$ field. Slight 
central peak in the curve of growth. Significant amounts of dust \citep{hugh03}.

\subsection{NGC~3021}

\begin{figure}
\epsscale{1.0}
\plotone{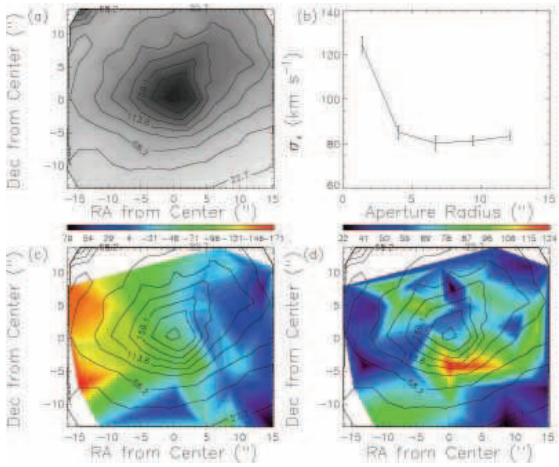}
\caption[NGC~3021]{
Same as Figure~\ref{fig:289} but for NGC~3021.
}\label{fig:3021}
\end{figure}

Figures~\ref{fig:3021} and \ref{fig:rot2964-3310}(b). Significant discordance and low S/N. Clear rotation and a peaked $\sigma_{\ast}$ 
field. Significant difference between extended and central apertures. 

\subsection{NGC~3162}

\begin{figure}
\epsscale{1.0}
\plotone{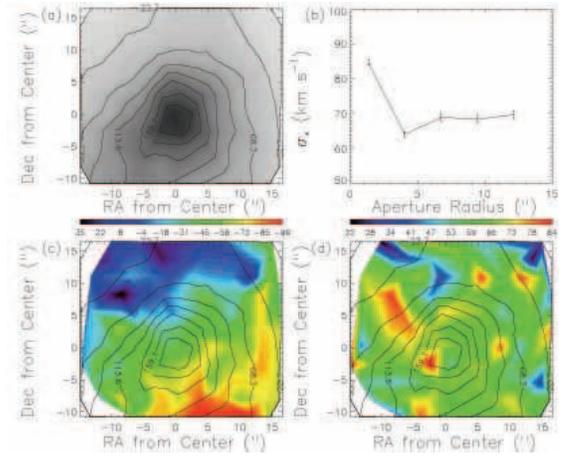}
\caption[NGC~3162]{
Same as Figure~\ref{fig:289} but for NGC~3162.
}\label{fig:3162}
\end{figure}

Figures~\ref{fig:3162} and \ref{fig:rot2964-3310}(c). Low S/N and typical discordance. Clear rotation and irregular $\sigma_{\ast}$ field. 
Significant difference between extended and central apertures. There are central asymmetries in the LOSVD. These are the first measures of 
$\sigma_{\ast}$ in NGC~3162.

\subsection{NGC~3310}

\begin{figure}
\epsscale{1.0}
\plotone{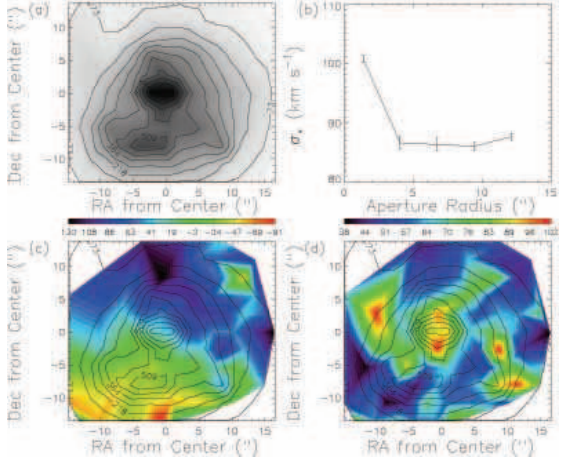}
\caption[NGC~3310]{
Same as Figure~\ref{fig:289} but for NGC~3310.
}\label{fig:3310}
\end{figure}

\begin{figure}
\epsscale{1.0}
\plotone{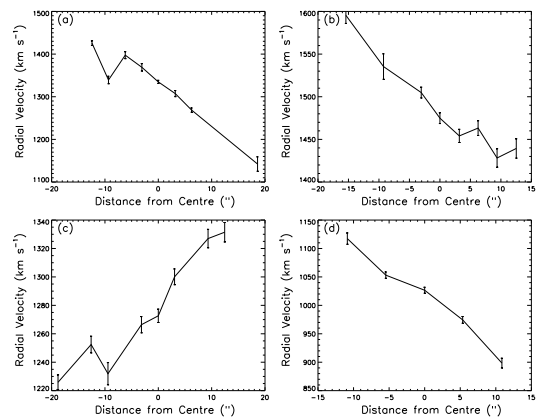}
\caption[Rotation Curves, NGC~2964...NGC~3310]{
Same and Figure~\ref{fig:rot613-2903} but for (a) NGC~2964 [90]. (b) NGC~3021 [90]. (c) NGC~3162 [30]. (d) NGC~3310 [0].
}\label{fig:rot2964-3310}
\end{figure}

Figures~\ref{fig:3310} and \ref{fig:rot2964-3310}(d). Good S/N and typical discordance. Clear rotation, patchy $\sigma_{\ast}$ field and a 
clear difference between central and extended $\sigma_{\ast}$. Possible high velocity components in central LOSVD. Significant amounts of 
dust \citep{hugh03}. 

\subsection{NGC~3949}

\begin{figure}
\epsscale{1.0}
\plotone{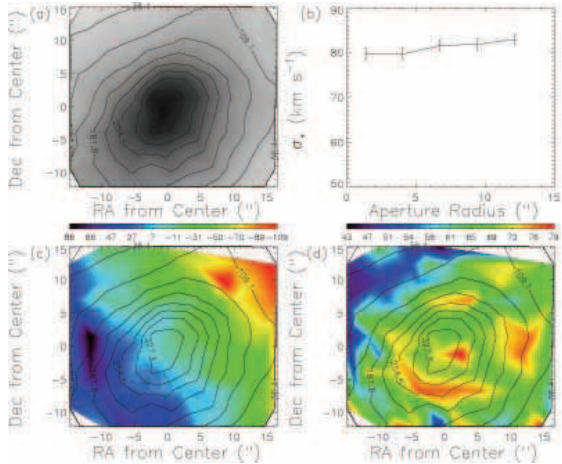}
\caption[NGC~3949]{
Same as Figure~\ref{fig:289} but for NGC~3949.
}\label{fig:3949}
\end{figure}

Figures~\ref{fig:3949} and \ref{fig:rot3949-4051}(a). High discordance and fair S/N. Clear rotation and irregular $\sigma_{\ast}$ field. 
Constant curve of growth. Significant high velocity components in the central LOSVD. Thin dust lane \citep{hugh03}. These are the first 
measures of $\sigma_{\ast}$ in NGC~3949.

\subsection{NGC~3953}

\begin{figure}
\epsscale{1.0}
\plotone{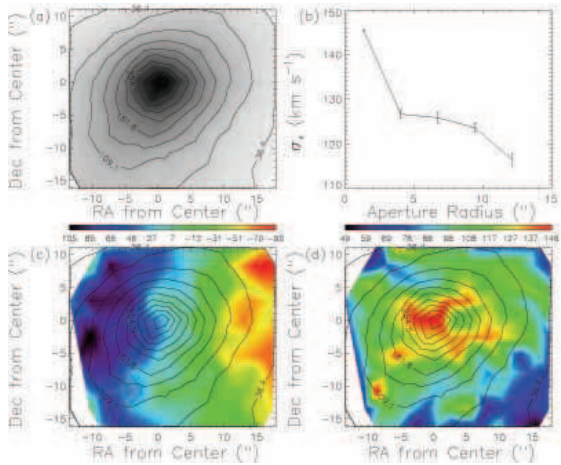}
\caption[NGC~3953]{
Same as Figure~\ref{fig:289} but for NGC~3953.
}\label{fig:3953}
\end{figure}

Figures~\ref{fig:3953} and \ref{fig:rot3949-4051}(b). Low discordance and good S/N. Centrally peaked $\sigma_{\ast}$ field and clear 
rotation. Large central curve of growth peak. There are no previous measures of $\sigma_{\ast}$ for NGC~3953. 

\subsection{NGC~4041}

\begin{figure}
\epsscale{1.0}
\plotone{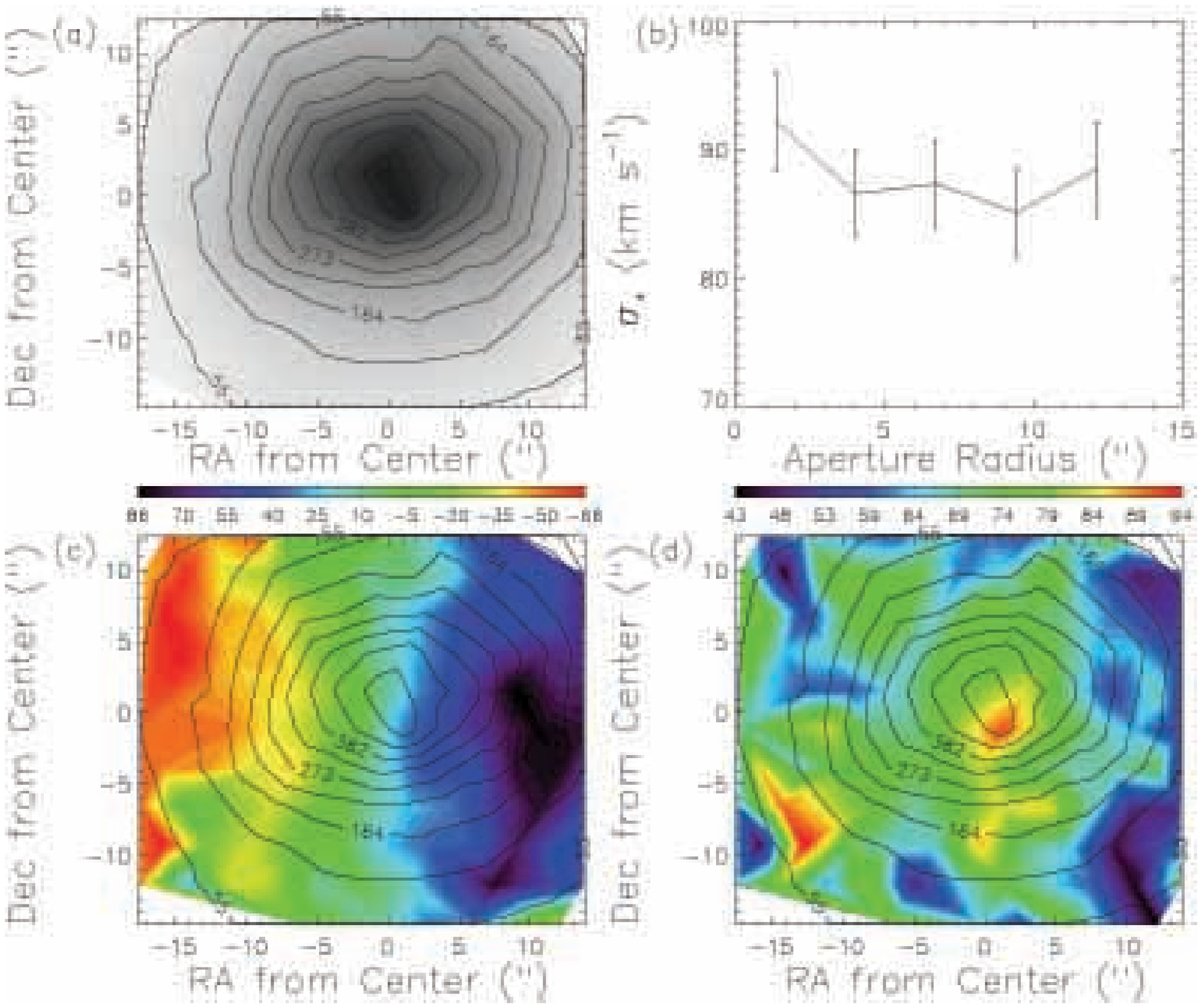}
\caption[NGC~4041]{
Same as Figure~\ref{fig:289} but for NGC~4041.
}\label{fig:4041}
\end{figure}

Figures~\ref{fig:4041} and \ref{fig:rot3949-4051}(c). Good S/N and low discordance. Fairly constant curve of growth, regular rotation and 
a centrally peaked $\sigma_{\ast}$ field. Nuclear star cluster \citep{mar03}. There are no previous measures of $\sigma_{\ast}$ for 
NGC~4041.

\subsection{NGC~4051}

\begin{figure}
\epsscale{1.0}
\plotone{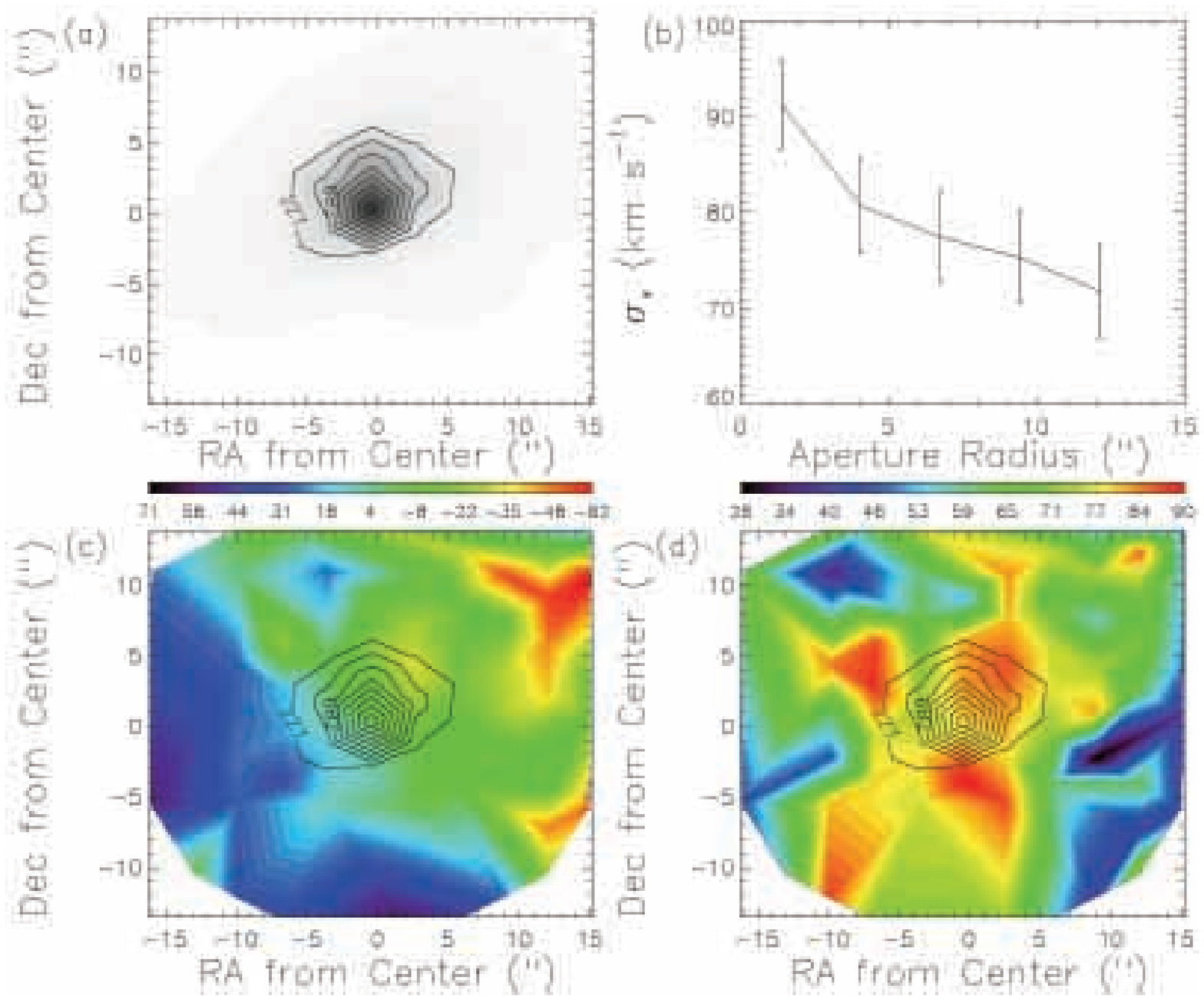}
\caption[NGC~4051]{
Same as Figure~\ref{fig:289} but for NGC~4051.
}\label{fig:4051}
\end{figure}

\begin{figure}
\epsscale{1.0}
\plotone{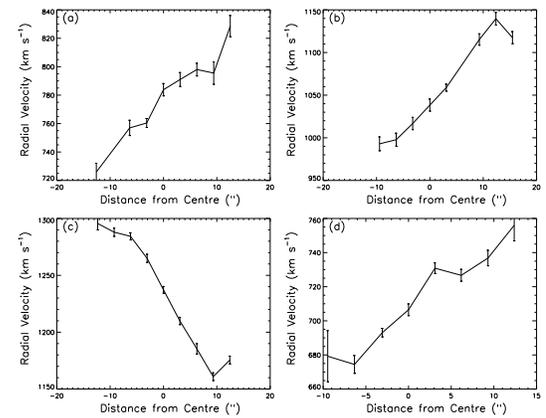}
\caption[Rotation Curves, NGC~3949...NGC~4051]{
Same as Figure~\ref{fig:rot613-2903} but for (a) NGC~3949 [120]. (b) NGC~3953 [90]. (c) NGC~4041 [90]. (d) NGC~4051 [120].
}\label{fig:rot3949-4051}
\end{figure}

Figures~\ref{fig:4051} and \ref{fig:rot3949-4051}(d). Significant discordance and low S/N. The large peak signal is due to the AGN. 
Regular rotation, patchy $\sigma_{\ast}$ field and a steep curve of growth. There are large values for both $h_3$ and $h_4$, the signature 
of a large asymmetry and super-Gaussian wings in the LOSVD.

\subsection{NGC~4088}

\begin{figure}
\epsscale{1.0}
\plotone{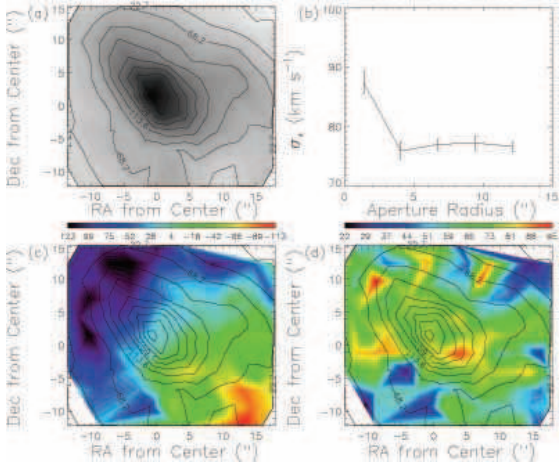}
\caption[NGC~4088]{
Same as Figure~\ref{fig:289} but for NGC~4088.
}\label{fig:4088}
\end{figure}

Figures~\ref{fig:4088} and \ref{fig:rot4088-4303}(a). Low S/N and high discordance. Clear rotation and a patchy $\sigma_{\ast}$ field. A 
significant central peak in the curve of growth. The inclination of NGC~4088 ($\sim71\degr$) suggests there may be some obscuration from 
the disk. These are the first measures of $\sigma_{\ast}$ in NGC~4088.

\subsection{NGC~4212}

\begin{figure}
\epsscale{1.0}
\plotone{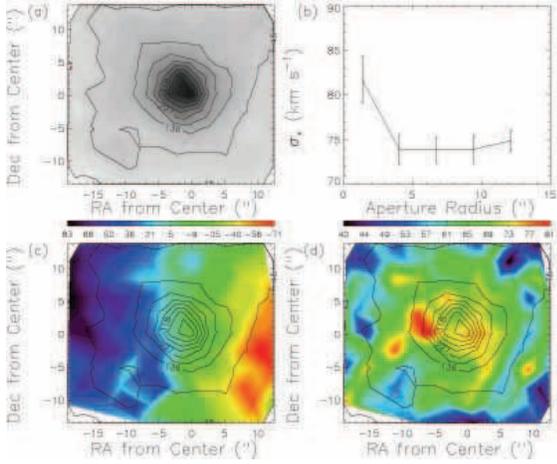}
\caption[NGC~4212]{
Same as Figure~\ref{fig:289} but for NGC~4212.
}\label{fig:4212}
\end{figure}

Figures~\ref{fig:4212} and \ref{fig:rot4088-4303}(b). Remarkably low discordance and good signal. Centrally peaked $\sigma_{\ast}$ field, 
clear regular rotation and a significant peak in the curve of growth. There are no previous measures of $\sigma_{\ast}$ for NGC~4212. 

\subsection{NGC~4258}

\begin{figure}
\epsscale{1.0}
\plotone{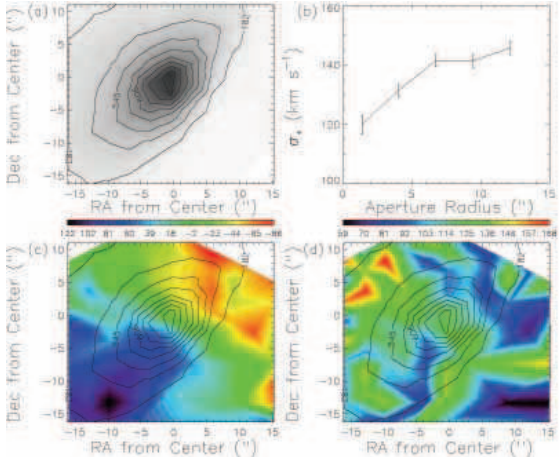}
\caption[NGC~4258]{
Same as Figure~\ref{fig:289} but for NGC~4258.
}\label{fig:4258}
\end{figure}

Figures~\ref{fig:4258} and \ref{fig:rot4088-4303}(c). Typical discordance and excellent S/N. Clearly defined rotation and an irregular 
$\sigma_{\ast}$ field. NGC~4258 is inclined at $\sim72\degr$ and exhibits a dust lane \citep{hugh03}. The curve of growth shows a steep 
drop toward the center.

\subsection{NGC~4303}

\begin{figure}
\epsscale{1.0}
\plotone{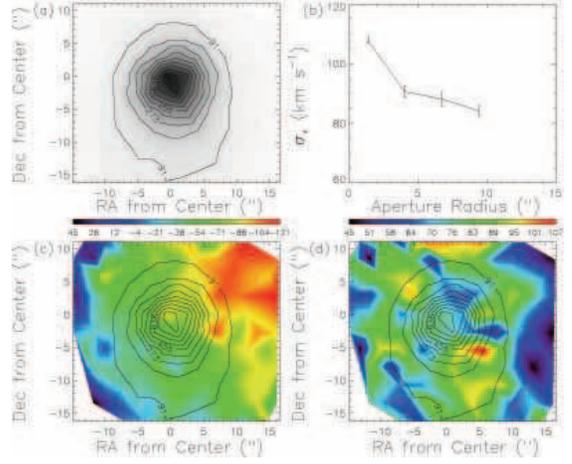}
\caption[NGC~4303]{
Same as Figure~\ref{fig:289} but for NGC~4303.
}\label{fig:4303}
\end{figure}

\begin{figure}
\epsscale{1.0}
\plotone{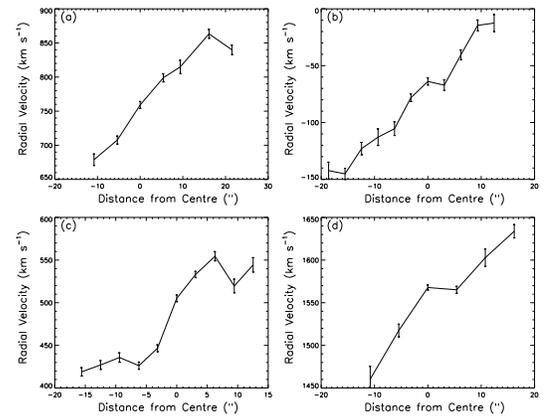}
\caption[Rotation Curves, NGC~4088...NGC~4303]{
Same as Figure~\ref{fig:rot613-2903} but for (a) NGC~4088 [30]. (b) NGC~4212 [90]. (c) NGC~4258 [120]. (d) NGC~4303 [120].
}\label{fig:rot4088-4303}
\end{figure}

Figures~\ref{fig:4303} and \ref{fig:rot4088-4303}(d). Good S/N and typical discordance. There is clear offset rotation and a patchy 
$\sigma_{\ast}$ field. Significant difference between extended and central $\sigma_{\ast}$. Nuclear spiral structure \citep{hugh03}.

\begin{figure}
\epsscale{1.0}
\plotone{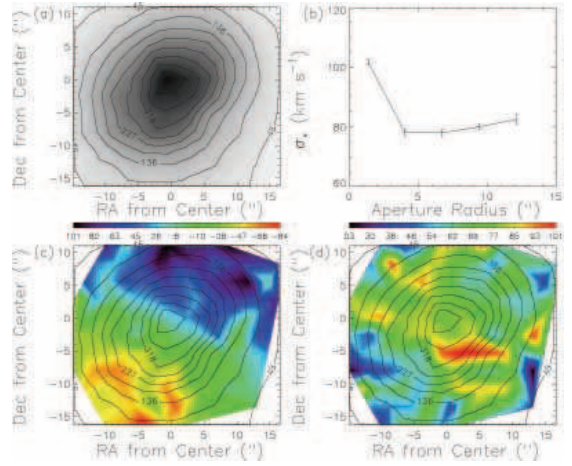}
\caption[NGC~4321]{
Same as Figure~\ref{fig:289} but for NGC~4321.
}\label{fig:4321}
\end{figure}

\subsection{NGC~4321}

Figures~\ref{fig:4321} and \ref{fig:rot4321-5055}(a). Low discordance and good S/N. Clear peak in $\sigma_{\ast}$ offset from the 
photometric center. Clear regular rotation and a centrally peaked curve of growth. 

\subsection{NGC~4536}

\begin{figure}
\epsscale{1.0}
\plotone{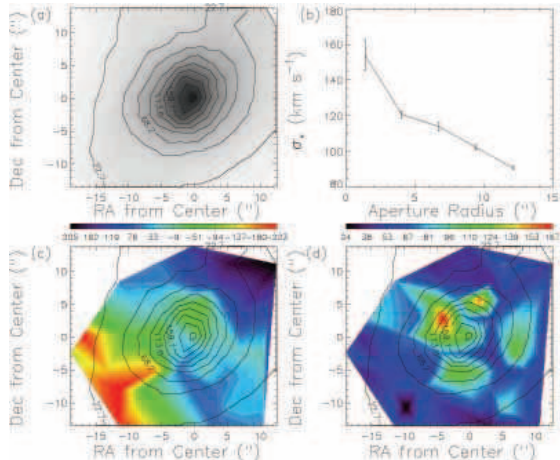}
\caption[NGC~4536]{
Same as Figure~\ref{fig:289} but for NGC~4536.
}\label{fig:4536}
\end{figure}

Figures~\ref{fig:4536} and \ref{fig:rot4321-5055}(b). Significant discordance and very low S/N. Steep rotation and centrally peaked 
$\sigma_{\ast}$ field. Steep centrally peaked curve of growth. High inclination ($\sim69\degr$) and obscuring dust present \citep{hugh03}.

\subsection{NGC~5005}

\begin{figure}
\epsscale{1.0}
\plotone{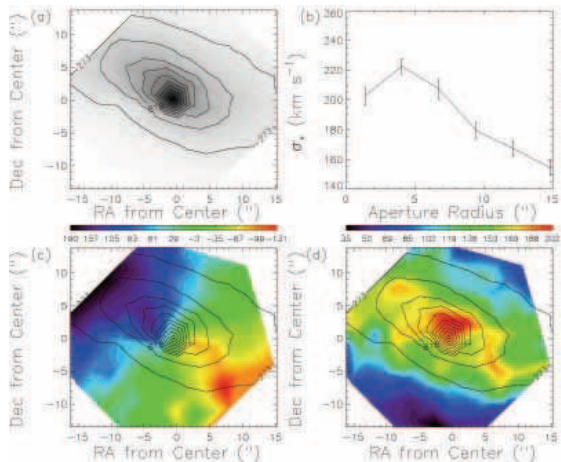}
\caption[NGC~5005]{
Same as Figure~\ref{fig:289} but for NGC~5005.
}\label{fig:5005}
\end{figure}

Figures~\ref{fig:5005} and \ref{fig:rot4321-5055}(c). Low discordance and excellent S/N. Clear, well defined rotation and complex 
$\sigma_{\ast}$ field. Highly variable curve of growth. Large inclination ($\sim67\degr$) and complicated dust morphology \citep{hugh03}. 
These are the first measures of $\sigma_{\ast}$ in NGC~5005.

\subsection{NGC~5055}

\begin{figure}
\epsscale{1.0}
\plotone{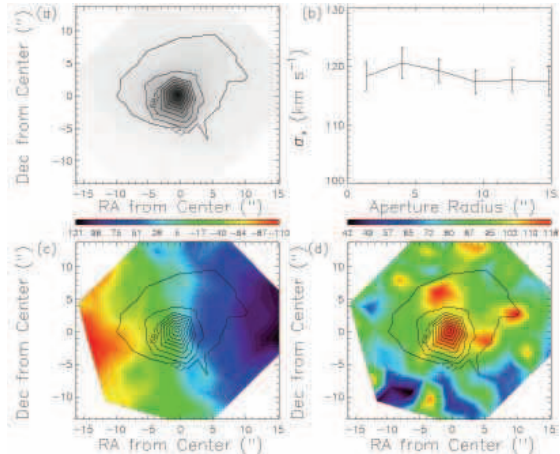}
\caption[NGC~5055]{
Same as Figure~\ref{fig:289} but for NGC~5055.
}\label{fig:5055}
\end{figure}

\begin{figure}
\epsscale{1.0}
\plotone{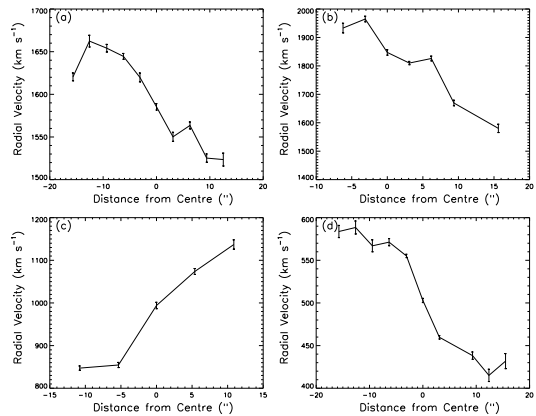}
\caption[Rotation Curves, NGC~4321...NGC~5055]{
Same as Figure~\ref{fig:rot613-2903} but for (a) NGC~4321 [120]. (b) NGC~4536 [120]. (c) NGC~5005 [30]. (d) NGC~5055 [90].
}\label{fig:rot4321-5055}
\end{figure}

Figures~\ref{fig:5055} and \ref{fig:rot4321-5055}(d). Low discordance and excellent S/N. Constant curve of growth and clear, well defined 
rotation. Patchy $\sigma_{\ast}$ field.

\section{Summary and Discussion}\label{summary}

In order to populate the low mass regimes of the SMBH - host relations we have completed a STIS survey of 54 bulges in spiral galaxies 
from which to derive the SMBH mass from gas kinematics \citep{mar03,atk04}. To place these points on the relations we need to determine 
the dynamical properties of the host bulges. To this end we have initiated the first IFS survey specifically directed toward spiral 
bulges. We have presented an atlas of 23 spiral bulges, eleven of which do not have previous estimates of $\sigma_{\ast}$. 

The properties of the LOSVDs were derived using the methods of Cross Correlation (Xcor) and Maximum Penalized Likelihood. Each method 
was thoroughly tested against a simulated system and are robust to spectral contamination from AGN and starbursts. 

Where the S/N has allowed we see that each system shows clear coherent radial rotation and patchy irregular velocity dispersion fields. 
This type of behaviour is displayed in a number of systems observed by SAURON, e.g. NGC~4382 (S0*(s)pec), NGC~4473 (E5) and NGC~4526 
(SABO*(s)) \citep{dez02,ems04}. We also note that the derived curves of growth, i.e. the variation of $\sigma_{\ast}$ with aperture 
size, show large variations of $\sigma_{\ast}$ in some cases. Generally we see offsets between the photometric centers of these systems 
and their peak values of $\sigma_{\ast}$. We are confident that this observation is also robust in the cases where a large amount of 
data has been excluded due to low S/N or template mismatching. This discordant data is mainly confined to the outer areas of the 
Integral Field Spectroscopy arrays where the S/N is at its lowest. As a consequence data at the centers of the arrays are still of high 
quality (S/N $> 14.5$). There is also a correlation between the complexity of $\sigma_{\ast}$ fields and the inclination of the system, 
especially where it is known that dust obscuration is significant, i.e., in high inclination cases ($\gtrsim70\degr$; NGC~2748, 
NGC~4088, NGC~4258, NGC5005) the larger values of $\sigma_{\ast}$ are confined to a band coincident with the photometric major axis. 
These observations suggest that $\sigma_{\ast}$ - the projected stellar velocity dispersion - is a non-trivial parameter to derive 
consistently from system to system.

With these considerations in mind we are left to contemplate the errors in SMBH mass estimates derived from the 
$M_{\bullet}-\sigma_{\ast}$ relation. The data presented here suggests that values of $\sigma_{\ast}$ taken from the literature, where 
little heed has been taken of aperture sizes and the effects of rotation, may demonstrate variations due to their heterogeneous nature. 
An attempt to quantify these errors, with respect to aperture size, can be made by calculating $\bar{\delta\sigma_{\ast}}$, i.e. the 
mean difference between the extended ($\sigma_{\rm ce}$) and central ($\sigma_{\rm cc}$) values of $\sigma_{\ast}$ as determined from 
Xcor. Indeed, if we are to assume that $\sigma_{\rm ce}$ and $\sigma_{\rm cc}$ are a good representation of $\sigma_{\ast}$ within $R_e$ 
and $R_e/8$ respectively ($R_e$ being the effective or half-light radius), then $\bar{\delta\sigma_{\ast}}$ may also go some way toward 
explaining the scatter in the $M_{\bullet}-\sigma_{\ast}$ relation, especially in the spiral region where the most weighting is given to 
the slope ($\alpha$).

From the data presented in Table~\ref{tab:results} we find $\bar{\delta\sigma_{\ast}}\approx20{\rm~km~s^{-1}}$. This figure implies that 
the slope of the $M_{\bullet}-\sigma_{\ast}$ relation will be affected by the aperture size used to measure $\sigma_{\ast}$. It then 
becomes clear that a formal unambiguous definition of $\sigma_{\ast}$ must be made, undoubtedly through the use of Integral Field 
Spectroscopy where the exact behaviour of $\sigma_{\ast}$ can be followed across two dimensions, before a confident estimate of the 
slope and scatter in the $M_{\bullet}-\sigma_{\ast}$ relation can be made. It must also be noted that carrying this value of 
$\bar{\delta\sigma_{\ast}}$ through the two forms of the $M_{\bullet}-\sigma_{\ast}$ relation, i.e., $\alpha = 4.72$ 
\citep{mandf01} and $\alpha = 4.02$ \citep{tre02} using $100-120{\rm~km~s^{-1}}$, leads to SMBH mass ranges of 
$0.49-1.17\times10^7M_{\odot}$ and $0.83-1.73\times10^7M_{\odot}$ respectively. Therefore, in future, authors wishing to use the 
$M_{\bullet}-\sigma_{\ast}$ relation to estimate the masses of SMBHs from ``simple'' $\sigma_{\ast}$ measurements must pay close 
attention to the processes used in order to make that estimate of $\sigma_{\ast}$ in the first place.

\acknowledgments

Partial support for this work was obtained from a PPARC studentship. We wish to thank the anonymous referee for contributing useful 
comments and suggestions that have improved the paper. The WHT is operated on the island of La Palma by the Isaac Newton Group in the 
Spanish Observatorio del Roque de los Muchachos of the Instituto de Astrofisica de Canarias. The AAT is operated by the AAO on behalf of 
the astronomical communities of Australia and the UK.

{\it Facilities}: \facility{ING:Herschel(IFU INTEGRAL)}, \facility{AAT(IFU SPIRAL)}

\clearpage


\begin{thebibliography}{}
\bibitem[{\rm Atkinson~et~al.(2005)}]{atk04}Atkinson, J. el~al. 2005, MNRAS in press, astroph/0502573
\bibitem[{\rm Alonso-Herrero, Ryder \& Knapen(2001)}]{ark01}Alonso-Herrero, A., Ryder, S. D. \& Knapen, J. H. 2001, MNRAS, 322, 757
\bibitem[{\rm Bender(1990)}]{bend90}Bender, R. 1990, A\&A, 229, 441
\bibitem[{\rm Cappellari \& Emsellem(2004)}]{capp04} Cappellari, M., \& Emsellem, E.\ 2004, \pasp, 116, 138 
\bibitem[{\rm de Zeeuw et~al.(2002)}]{dez02}de Zeeuw, P.~T., et al. 2002, \mnras, 329, 513
\bibitem[{\rm Dressler(1984)}]{dre84}Dressler, A. 1984, ApJ, 286, 97
\bibitem[{\rm Emsellem~et~al.(2004)}]{ems04}Emsellem, E. et al. 2004, MNRAS, 352, 721
\bibitem[{\rm Ferrarese \& Merritt(2000)}]{fandm00} Ferrarese, L.~\& Merritt, D.\ 2000, \apjl, 539, L9 
\bibitem[{\rm Franx, Illingworth \& Heckman(1989)}]{fih89}Franx, M., Illingworth, G. \& Heckman, T. 1989, ApJ, 344, 613
\bibitem[{\rm Gebhardt~et~al.(2000)}]{geb00}Gebhardt et al. 2000, ApJ, 539, L13
\bibitem[{\rm Graham et~al.(2003)}]{gra03}Graham, A. W., Erwin, P., Caon, N. \&  Trujillo, I. 2003, RevMexAA, 17, 196
\bibitem[{\rm Hughes~et~al.(2003)}]{hugh03}Hughes, M. A. et al. 2003, AJ, 126, 742
\bibitem[{\rm Kormendy \& Richstone(1995)}]{kandr95}Kormendy, J. \& Richstone, D. O. 1995, ARAA, 33, 581
\bibitem[{\rm Kuijken~\&~Merrifield(1993)}]{kandm93}Kuijken, K. \& Merrifield, M. R. 1993, MNRAS, 264, 712
\bibitem[{\rm Kurtz~\&~Mink(1998)}]{kandm98}Kurtz, M. J \& Mink, D. J. 1998, PASP, 110, 934
\bibitem[{\rm Laor(2001)}]{laor01}Laor, A. 2001, ApJ, 553, 677
\bibitem[{\rm Magorrian~et~al.(1998)}]{mag98}Magorrian, J. et al. 1998, AJ, 115, 2285
\bibitem[{\rm Marconi \& Hunt(2003)}]{mandh03}Marconi, A. \& Hunt, L. K. 2003, ApJL, 589, L21
\bibitem[{\rm Marconi~et~al.(2003)}]{mar03}Marconi, A. et al. 2003, ApJ, 586, 868
\bibitem[{\rm Merritt(1997)}]{merr97}Merritt, D. 1997, AJ, 114, 228
\bibitem[{\rm Merritt \& Ferrarese(2001)}]{mandf01}Merritt, D. \& Ferrarese. L. 2001, ApJ, 547, 140
\bibitem[{\rm Rix~\&~White(1992)}]{randw92}Rix, H. \& White, S. D. M. 1992, MNRAS, 254, 389
\bibitem[{\rm Sargent et~al.(1977)}]{sarg77}Sargent, W. L. W., Schechter, P. L., 
        Boksenberg, A. \& Shortridge, K. 1977, ApJ, 212, 326
\bibitem[{\rm Scarlata~et~al.(2004)}]{scar04}Scarlata, C., et al. 2004, \aj, 128, 1124 
\bibitem[{\rm Simkin(1974)}]{sim74}Simkin, S. M. 1974, A\&A, 31, 129
\bibitem[{\rm Tonry~\&~ Davies (1979)}]{tandd79}Tonry, J. \& Davies, M. 1979, AJ, 84, 1511
\bibitem[{\rm Tremaine et~al.(2002)}]{tre02}Tremaine, S., et al. 2002, ApJ, 574, 740
\bibitem[{\rm van~der~Marel~\&~Franx(1993)}]{vdmandf93}van der Marel, R. P. \& Franx, M. 1993, ApJ, 407, 525
\end{thebibliography}
\end{document}